# Contaminants removal and bacterial activity enhancement along the flow path of constructed wetland microbial fuel cells

Marco Hartl[a,b], Diego F. Bedoya-Ríos[c], Marta Fernández Gatell[a], Diederik P.L. Rousseau[b], Gijs Du Laing[b], Marianna Garfí[a], Jaume Puigagut[a,*]

[a] GEMMA - Environmental Engineering and Microbiology Research Group, Department of Civil and Environmental Engineering, Universitat Politècnica de Catalunya·BarcelonaTech, c/ Jordi Girona 1-3, Building D1, E-08034 Barcelona, Spain.

[b] Department of Green Chemistry and Technology, Faculty of Bioscience Engineering, Ghent University. Coupure Links 653, 9000 Gent, Belgium.

[c] Grupo Ciencia e Ingeniería del Agua y el Ambiente, Facultad de Ingeniería, Pontificia Universidad Javeriana - Bogotá D.C.- Carrera 7 No. 40 – 62, Colombia.

*Corresponding author:
Tel: +34 93 401 08 98
Fax: +34 93 401 73 57
Email: Jaume.Puigagut@upc.edu

**Abstract**

Microbial fuel cells implemented in constructed wetlands (CW-MFCs), albeit a relatively new technology still under study, have shown to improve treatment efficiency of urban wastewater. So far the vast majority of CW-MFC systems investigated were designed as lab-scale systems working under rather unrealistic hydraulic conditions using synthetic wastewater. The main objective of this work was to quantify CW-MFCs performance operated under different conditions in a more realistic setup using meso-scale systems with horizontal flow fed with real urban wastewater. Operational conditions tested were organic loading rate (4.9±1.6, 6.7±1.4 and 13.6±3.2 g COD/$m^2$.day) and hydraulic regime (continuous vs intermittent feeding) as well as different electrical connections: CW control (conventional CW without electrodes), open-circuit CW-MFC (external circuit between anode and cathode not connected) and closed-circuit CW-MFC (external circuit connected).

Eight horizontal subsurface flow CWs were operated for about four months. Each wetland consisted of a PVC reservoir of 0.193 $m^2$ filled with 4/8 mm granitic riverine gravel (wetted depth 25 cm). All wetlands had intermediate sampling points for gravel and interstitial liquid sampling. The CW-MFCs were designed as three MFCs incorporated one after the other along the flow path of the CWs. Anodes consisted of gravel with an incorporated current collector (stainless steel mesh) and the cathode consisted of a graphite felt layer. Electrodes of closed-circuit CW-MFC systems were connected externally over a 220 Ω resistance.

Results showed no significant differences between tested organic loading rates, hydraulic regimes or electrical connections, however, on average, systems operated in closed-circuit CW-MFC mode under continuous flow outperformed the other experimental conditions. Closed-circuit CW-MFC compared to conventional CW control systems showed around 5% and 22% higher COD and ammonium removal, respectively. Correspondingly, overall bacteria activity, as measured by the fluorescein diacetate technique, was higher (4% to 34%) in closed-circuit systems when compared to CW control systems.

1. **INTRODUCTION**

Constructed wetlands (CWs) are engineered systems for water and wastewater treatment, simulating processes occurring in nature (Vymazal, 2011). Treatment in CWs is based on physical, chemical and biological processes. The treatment beds consist of shallow lined basins filled with a filter media (generally gravel or sand) and are commonly planted with aquatic macrophytes (García et al., 2010). CWs treat wastewater from a wide range of sources, such as domestic, industrial and agricultural wastewater or landfill leachate, in different climate zones around the world (Langergraber and Haberl, 2001; Molle et al., 2005). These natural systems are characterized by their low energy demand, comparative low cost, easy operation and maintenance as well as the possibility to use local materials and labor for their construction. Hence, they have a strong potential for application as an alternative to conventional systems for sanitation of small communities, also in rural areas and emerging countries (García, 2001; Kivaisi, 2001; Puigagut et al., 2007). A disadvantage of CWs is their relatively high area demand of ca. 1-10 $m^2$/p.e. (Kadlec and Wallace, 2009).

Microbial Fuel Cells (MFCs) are bioelectrochemical systems that generate current by means of electrochemically active microorganisms as catalysts (Logan et al., 2006). In a MFC, organic and inorganic substrates are oxidized by bacteria and the electrons are transferred to the anode from where they flow through a conductive material and a resistor to an electron acceptor, such as oxygen, at the cathode (Logan et al., 2006; Rabaey et al., 2007). Compounds oxidized at the anode are mainly simple carbohydrates such as glucose or acetate that can be already present in the environment or obtained from the microbial degradation of complex organic substrates such as organic sediments or wastewater (Min and Logan, 2004; Reimers et al., 2001). Therefore, MFCs are able to harvest energy in the form of electricity directly from wastewater (Du et al., 2007; Lefebvre et al., 2011; Min and Logan, 2004).

MFC systems can exploit the naturally occurring redox gradient in horizontal subsurface flow (HF) CWs. The first publication on CWs incorporating MFCs (CW-MFCs) appeared in 2012 and was published by Yadav et al. (2012). Since then publications on the subject per year are increasing, resulting in a rough total of around 79 up until March 2018.

So far the vast majority of CW-MFC systems investigated are designed as lab-scale systems working under rather unrealistic hydraulic conditions (up-flow, batch feeding) using synthetic wastewater (Corbella et al., 2016b; Doherty et al., 2015; Fang et al., 2016; Liu et al., 2012; Oon et al., 2017; Song et al., 2017; Srivastava et al., 2015; Villaseñor et al., 2013; Wang et al., 2017; F. Xu et al., 2018; Xu et al., 2017; Zhao et al., 2013).

As indicated above, the implementation of MFCs in CWs is a relatively new research field, and current available information on this topic is mostly focused on optimizing treatment efficiency and energy production. Conventional MFCs are able to produce up to 12 $W·m^{-3}$ electricity (Logan and Rabaey, 2012). However, due to high internal resistances the highest reported electrical output from CW-MFCs is 2 $W·m^{-3}$ (Xu et al., 2017), whereas averages for most systems are even a magnitude lower. Systems using wastewater reported electricity production of 9.4 $mW·m^{-2}$ (Zhao et al., 2013) and 276 $mW·m^{-3}$ (Doherty et al., 2015). In comparison to solar panels with for example 175 $W/m^2$ (Panasonic HIT® Photovoltaic Module, 2012) it seems that electricity production alone from wastewater by MFC or CW-MFC technology is currently not a feasible goal.

Besides energy production, CW-MFC systems can also improve the treatment of organic matter. When comparing closed-circuit (MFC anode and cathode externally connected) and open-circuit (MFC anode and cathode externally not connected) lab-scale results, Katuri et al. (2011) showed 16-20% higher COD removal for closed-circuit MFC systems. The same tendency was observed by Srivastava et al. (2015) with 16-20% higher COD removal in closed-circuit compared to open-circuit CW-MFCs and even 10-31% higher performance compared to conventional CWs (without anode and cathode). Exemplary COD removal efficiencies in CW-MFC are 75% (Yadav et al., 2012), 82% (Xu et al., 2018), 76.5% (Zhao et al., 2013) and even up to 100% (Oon et al., 2015), however the latter used artificial aeration. As mentioned before, most of the systems investigated so far do not reproduce realistic HF CW conditions due to the flow direction and geometry of systems (often up-flow in tubular reactors), and smaller internal resistances than in full-scale implementation due to smaller distances between electrodes and other factors. In general the presence of an insoluble electron acceptor, i.e. an anode, showed to increase the metabolic rate of anaerobic bacteria (Fang et al., 2013) and seems to be a beneficial environment for the growth of bacteria apart from electrogens as well; Xu et al. (2018) found that the microbial community´s richness and diversity is higher in closed-circuit systems and also Wang et al. (2016b) found higher richness in closed-circuit as compared to open-circuit CW-MFC systems. Additionally, electroactive bacteria seem to outperform other microbial communities (Zhang et al., 2015).

Apart from organic matter, MFC studies have shown that closed-circuit MFCs show a higher ammonium treatment efficiency than open-circuit MFCs (Kim et al., 2008; Lu et al., 2009). This increased ammonium removal efficiency could also be observed in CW-MFCs by Corbella and Puigagut (2018) with ammonium removal efficiencies of 66±14% and 53±17% for closed-circuit and open-circuit mode, respectively.

The main objective of this work was to quantify and improve the treatment efficiency of urban wastewater with CW-MFCs. The effect of hydraulic regime (continuous/intermittent) and organic loading rate (4.9±1.6, 6.7±1.4 and 13.6±3.2 g COD/$m^2$.day) on CW-MFCs performance and the effect of CW-MFCs on bacterial activity along the flow path of the treatment bed are also discussed. The authors believe that this work will provide a useful insight into the actual net contribution of CW-MFCs on the treatment of urban wastewater. In spite of the lack of plants in the systems, the CW-MFCs used in this research could give additional information on the pollutant removal in larger scale systems under more realistic CWs design and operation conditions; also the here used configuration with three MFCs incorporated one after the other along the flow path of the CWs and the associated measured current along the flow path together with the measured bacterial activity will help to provide a better insight into the bioelectrochemical behavior and nutrient removal of CW-MFCs.

## 2. MATERIALS AND METHODS

### 2.1 General design

For the purpose of this work, eight meso-scale horizontal subsurface flow (HF) CW-MFC systems consisting of a PVC reservoir of ca. 0.193 $m^2$ (55 x 35 cm) surface area filled up with 4/8 mm granitic riverine gravel were constructed. The systems were not planted in order to not add another influencing parameter and further increase the experiment complexity. Campaigns with planted CW-MFC duplicates are planned for the future. Wetted depth was set to be 25 cm. At the inlet and around the drainage of the outlet 7/14 mm granitic riverine gravel was used.

The CW-MFCs were designed as three MFCs incorporated one after the other along the flow path of the CWs. Therefore, the experimental systems were operated as a three-MFC system (see Figure 1). Each electrode consisted of an anode with four stainless steel mesh rectangles (Figure 1, C) (SS marine grade A316L, mesh width=4.60 mm, Øwire=1.000 mm, S/ISO 9044:1999) in series (4 cm away from each other). Each metal mesh covered nearly the whole cross-sectional area (0.08 $m^2$) of the CW. Each cathode consisted of a carbon felt mat (Figure 1, D) (1.27 cm thick, with a projected surface of 0.03 $m^2$, 99.0% carbon purity). A layer of glass wool was placed underneath the cathodes in order to avoid any oxygen leaking from the cathode down to the anode as recommended elsewhere (Venkata Mohan et al., 2008). For the connected systems (closed-circuit), each electrode´s anode and cathode were externally connected via a 220 Ω resistance, selected according to results by Corbella and Puigagut (2018). The voltage across the external resistance for each electrode was continuously monitored by means of a datalogger (Campbell Scientific CR1000, AM16/32B Multiplexor). For the open-circuit systems, the anode and cathode were not connected (open-circuit). For the conventional HF CW control (operated from week 12 to week 23), metal meshes were removed from two of the systems that were previously operated under open-circuit conditions.

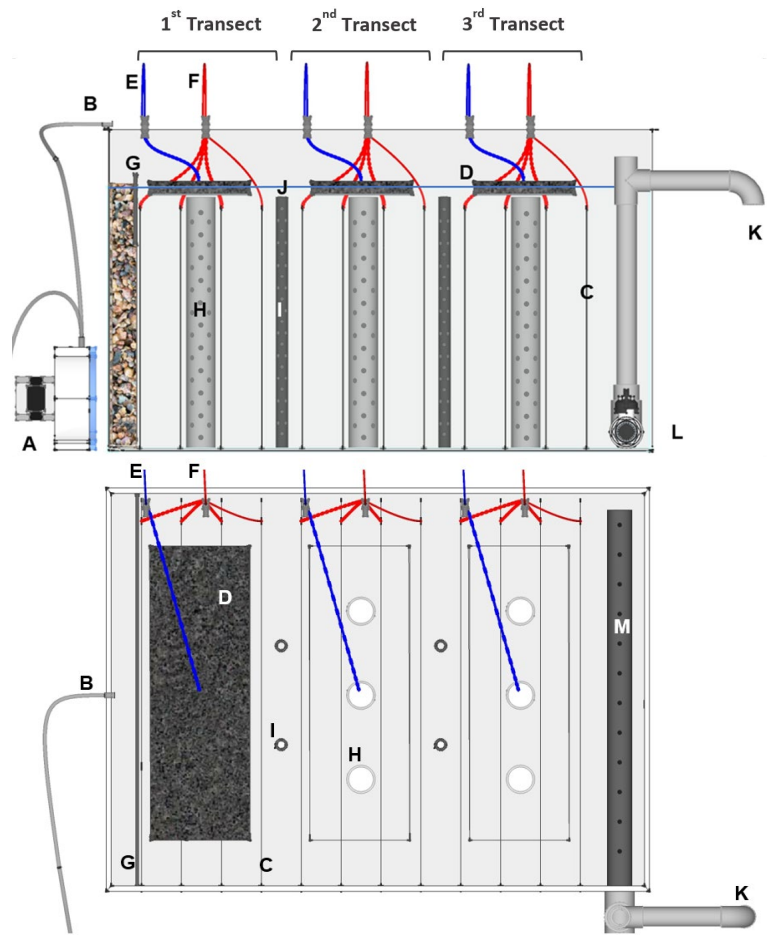

**Figure 1. Section- (top) and plan-view (bottom) of the CW-MFC systems. A: Pump; B: Inflow; C: Anode; D: Cathode; E/F: Anode/Cathode connection to datalogger; G: Inflow barrier to avoid water short-circuiting on surface; H: Gravel core sampling tubes; I: Liquid sampling tubes; J: Water level; K: Standing pipe effluent; L: Drainage; M: Effluent collection tube.**

Intermediate liquid sampling ports were installed after the first third and second third (Figure 1, I), separating the first, second and third transect of the systems which are basically congruent with the three successive MFCs of the wetland. These sampling ports consisted of two perforated plastic tubes (Ø=1cm, positioned vertically 5 cm left and right of center). Underneath each cathode three perforated plastic tubes (Ø=3.2cm, positioned at the center and 8.5 cm left and right of the center) were placed and filled with a plastic mesh "sock" containing the same gravel material as the systems (Figure 1, H). These socks were removable and were used to test the bacterial activity along the flow path of the wetland.

## 2.2 Operational conditions

All systems received the same primary treated urban wastewater throughout the whole experimentation period (23 weeks within the period from May until December 2017 excluding breaks of 8 weeks during summer and the first week of December). Wastewater feeding started already 6 weeks before the start of experimentation in order to establish the biofilm in the systems. The wastewater was stored within a reservoir of ca. 180 L that was refilled every weekday in order to keep the organic matter concentration as stable as possible. Sampling and analysis were conducted once a week.

During the first 10 experimentation weeks (from May to July 2017) the effect of hydraulic regime and organic loading rate on the treatment performance of closed- and open-circuit systems was tested. The compared hydraulic regimes were continuous and intermittent feeding. Continuous flow mode systems received the same flow rate all day long, whereas intermittent flow systems received alternating 4 hours of double flow and 4 hours of no flow, resulting in the same total flow as continuous flow systems on a daily basis. The inflow was provided by peristaltic pumps (Damova MP-3035-6M) controlled by variable frequency drives (VFDs) (Toshiba VF-nC3S).

Two different hydraulic loading rates were applied, i.e. 26 and 52 mm/d. The higher rate was obtained by doubling the flow rate (and thereby dividing the HRT in half) resulting in a theoretical HRT and average organic loading rate (OLR) of 3.9±0.2 and 1.9±0.1 days and around 4.9±1.6 and 13.6±3.2 g COD/m².day, during low and high loading periods, respectively (the high OLR is not exactly the double of the low OLR due to natural variations of the urban wastewater used). During experimentation week 1-5 the eight systems were operated under low OLR, and during experimentation week 6-10 with high OLR. The parameter OLR was chosen over HRT for comparison of the periods due to the higher reliability in the calculation of the OLR as opposed to the HRT which is only a theoretical value and could be different to the real HRT in the systems. The other two factors of continuous/intermittent feeding and closed-/open-circuit electrical connection led to duplicates of each combination in the first 10 weeks of experimentation (see Table 1).

**Table 1.** Operational conditions during the 23 weeks of experimentation concerning organic loading, hydraulic regime and electrical connection within the systems as well as the resulting individual experimental setups of the eight systems

| Experi-mentation Week[a] | Organic loading rate (g COD/m².day) | Hydraulic regime | Electrical connection | Resulting system setup |
|---|---|---|---|---|
| 1-5 | Low OLR1 4.9±1.6 | Continuous or Intermittent | Closed-circuit or Open-circuit | 2x continuous flow / closed-circuit 2x continuous flow / open-circuit 2x intermittent flow / closed-circuit 2x intermittent flow / open-circuit |
| 6-10 | High OLR 13.6±3.2 | | | |
| 11 | Low OLR2 6.7±1.4 | Continuous | | 4x closed-circuit 4x open-circuit |
| 12-23 | | Continuous | Closed-circuit, Open-circuit or CW control | 4x closed-circuit 2x open-circuit 2x CW control |

[a] only weeks in which experiments were conducted, i.e. excl. 8 weeks during summer and first week of December

Starting from experimentation week 11 (in September 2017, after 6 weeks of summer break during which the systems were fed with water and two weeks of wastewater feeding to restart systems), the treatment efficiency experiments were continued (until end of December 2017, except for the first week of December), this time only with continuous flow and low HLR (ca. 26 mm/d) resulting in a theoretical HRT of 3.8±0.3 days and an average OLR of 6.7±1.4 g COD/m².day. Starting from experimentation week 12 two of the open-circuit CW-MFCs were converted to conventional HF CWs by removing the SS mesh anodes, creating a conventional CW control duplicate without electrodes, and still leaving two open-circuit CW-MFCs and four closed-circuit CW-MFCs for investigation on solely the impact of the different electrical connections for the remaining experimentation weeks 12-23 (see Table 1).

**2.3   Sampling and analysis**

Samples were taken weekly from the influent, the intermediate sampling points placed at 1/3 and 2/3 of the wetland length and the effluent of each system. Influent and effluent samples were grab samples collected from inlet and effluent tubes, respectively. Intermediate samples were 60 mL composite grab samples (four times 15 mL) extracted from the pairs of sampling tubes placed after 1/3 and 2/3 from the inlet by means of a syringe. From each tube, two samples were taken, at 15 and 5 cm depth (i.e., 10 and 20 cm from the bottom of the system). The parameters total chemical oxygen demand (COD), ammonium -N, nitrate -N, nitrite -N, sulfate and orthophosphate -P as well as total suspended solids (TSS) and volatile suspended solids (VSS) were analyzed according to standard methods (APHA-AWWA-WEF, 2005). Physical parameters such as wastewater temperature, dissolved oxygen (DO) concentration (both; EUTECH instruments, EcoScan DO 6) and pH (CRISON pH/mV – meter 506) were measured as well using portable devices. Statistical analysis was conducted using Kruskal-Wallis and Shapiro-Wilk tests as well as single-factor and two-factor (with replication) analysis of variance (ANOVA).

**2.4   Microbial activity analysis**

Microbial activity was determined by means of the fluorescein diacetate (FDA) hydrolysis, a technique that has shown to correlate well with microbial population and its activity (Adam and Duncan, 2001). The FDA is a colorless compound which can be hydrolyzed by different enzymes releasing fluorescein as an end product, which absorbs strongly at 490 nm. For this procedure, two (out of the four available) closed-circuit systems and two CW control systems were investigated, using the gravel cores contained within the sampling tubes located in each of the transects of the systems (see Figure 1). These gravel cores (three for each transect at a time) were introduced into previously constructed reactors of 10 cm diameter and 28 cm height (see Figure 2).

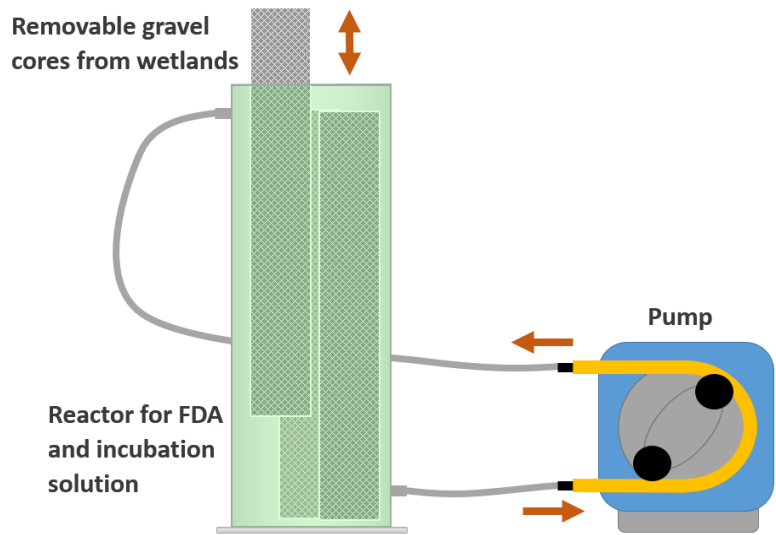

**Figure 2.** Microbial activity analysis setup including a reactor for the FDA and incubation solution in which the removable gravel cores (three per transect) from the wetland systems are submerged. The solution is mixed by means of a peristaltic pump.

At the time the three gravel cores were submerged the reactor already contained a prepared phosphate buffer at pH 7.6 together with one mL of 0.4 mM FDA (Acros Organics) resulting in a final concentration of $8 \cdot 10^{-4}$ mM FDA, following a similar but modified procedure by (Iasur-Kruh et al., 2010). This solution was recirculated with a pump and after 50 min a 2 ml sample was taken from the top of the reactor. Fluorescein released was measured using a spectrophotometer (Spectronic GENESYS 8 Thermo Scientific ™) at a wavelength of 490 nm and then converted to Fluorescein molar mass via a calibration curve. For the purpose of this study the final Fluorescein molar mass value is then called the microbial activity. Statistical analyses were conducted using Kruskal-Wallis and Shapiro-Wilk tests as well as single-factor ANOVA.

## 3. RESULTS AND DISCUSSION

### 3.1 Assessment of operational conditions to optimize CW-MFC along the flow path

#### 3.1.1 Overview

Table 2 shows an overview for COD, ammonium, nitrate, nitrite and orthophosphate removal results from inlet to outlet, expressed in total specific mass (g/m$^2$.d) for open-circuit and closed-circuit CW-MFC systems (see annex Table 4 for removal in percentage). Results are further divided into the three different OLR periods (low OLR 1 in first 5 weeks, high OLR in the following 5 weeks and low OLR 2 in the remaining 13 weeks) and different hydraulic regimes (continuous/intermittent) for low OLR 1 and high OLR period and only continuous flow in low OLR 2.

**Table 2.** COD, ammonium, nitrate, nitrite and orthophosphate average mass removal rate (g/m².d) with standard deviation from inlet to outlet for low OLR 1, high OLR and low OLR 2 as well as intermittent or continuous flow hydraulic regime for open-circuit (OC) and closed-circuit (CC) CW-MFC systems

| Removal (g/m$^2$.d) | | Low OLR 1 (week 1-5) 4.9±1.6 g COD/m².day | | High OLR (week 6-10) 13.6±3.2 g COD/m².day | | Low OLR 2[a] (week 11-23) 6.7±1.4 g COD/m².day |
|---|---|---|---|---|---|---|
| | | Intermittent flow | Continuous flow | Intermittent flow | Continuous flow | Continuous flow |
| COD (n=4/5/11)[b] | OC | 3.0±1.6 | 3.0±1.8 | 8.3±3.5 | 8.5±3.7 | 4.6±1.0 |
| | CC | 2.8±1.7 | 3.0±1.8 | 9.6±3.9 | 9.6±2.9 | 4.9±1.1 |
| NH$_4$ -N (n=4/5/7)[b] | OC | 0.2±0.1 | 0.2±0.1 | 0.5±0.7 | 0.6±0.6 | 0.3±0.2 |
| | CC | 0.2±0.1 | 0.3±0.1 | 0.7±0.5 | 0.8±0.4 | 0.5±0.3 |
| NO$_3$ -N (n=4/4/8)[b] | OC | -0.009±0.026 | -0.013±0.061 | 0.005±0.014 | -0.002±0.018 | 0.000±0.000 |
| | CC | -0.012±0.035 | -0.032±0.064 | -0.022±0.033 | -0.065±0.042 | -0.011±0.012 |
| NO$_2$ -N (n=4/4/8)[b] | OC | 0.023±0.052 | 0.039±0.078 | 0.094±0.235 | -0.075±0.125 | -0.004±0.014 |
| | CC | 0.028±0.058 | 0.058±0.080 | 0.057±0.114 | -0.154±0.046 | -0.002±0.020 |
| PO$_4$ -P (n=4/4/8)[b] | OC | 0.02±0.03 | 0.03±0.01 | 0.03±0.04 | 0.03±0.04 | 0.01±0.01 |
| | CC | 0.02±0.02 | 0.03±0.02 | 0.02±0.04 | 0.04±0.06 | 0.01±0.03 |

[a] Low OLR 2 results are shown in more detail in section 3.2 on the electrical connection effects
[b] Some experimentation weeks could not be considered due to highly diluted influent or technical analysis problems

With regards to different organic loading periods, only continuously fed systems are discussed and compared for all nutrients, since COD and ammonium treatment, though not being significantly different, were generally higher in continuously fed systems. In addition, continuously fed systems showed a very significant higher current density generation within the first transect (see Figure 3).

### 3.1.2 Hydraulic regime effects

In general, closed-circuit and continuously fed systems tended to show higher nutrient removal efficiencies when compared to the rest of operational conditions tested, although no statistically significant differences in COD or ammonium removal were found (for details see annex Table 5). When comparing different hydraulic regimes with the same electrical connection, closed-circuit continuous systems had only 2 and 1% higher COD removal than closed-circuit intermittent systems during low OLR 1 and high OLR period, respectively. Open-circuit continuous systems had 2% lower and 4% higher COD removal than open-circuit intermittent systems during low OLR 1 and high OLR period, respectively. As expected, the majority of COD was removed within the first transect, since organic matter removal basically follows a first-order degradation (Kadlec and Wallace, 2009).

Ammonium removal rates did not show any significant differences between hydraulic regimes and electrical connections (for details see annex Table 5) but exhibited the same tendency as COD but more pronounced, with continuously fed and closed-circuit systems showing higher removal rates. When comparing different hydraulic regimes within the same electrical connection, closed-circuit continuous systems showed, in average, 11% and 4% higher ammonium removal than closed-circuit intermittent systems during low OLR 1 and high OLR period, respectively. Open-circuit continuous systems had 6 and 12% higher ammonium removal than open-circuit intermittent systems during low OLR 1 and high OLR period, respectively.

Continuously fed systems tended to have a higher nitrate increase throughout all OLR periods, with (an extremely) significant difference only in the high OLR period, probably caused by the shortened HRT (for details see annex Table 5). Continuously fed systems showed higher nitrite removal during low OLR 1 but also nitrite increase in these systems was higher during high OLR, however, without a significant difference (a significant difference was only found in terms of electric connection, for details see annex Table 5). The strong nitrite increase in continuously fed systems in the high OLR period could be a sign of a lack of oxygen and incomplete nitrification. Dissolved oxygen concentrations in the water column (3 cm and lower below water level) were below the detection limit of the probe along the whole flow path, i.e. at the inflow as well as after first, second and last transect.

An explanation for the slightly higher COD and ammonium removal in closed-circuit systems could be that continuous as compared to intermittent flow in HF CWs increases the vertical redox gradient and thereby provides a higher potential to drive MFC reactions (Corbella et al., 2014). The insignificance of differences could be partly due to the relatively high standard deviation, most likely caused by the variation in quality of the used real urban wastewater due to natural causes like rainfall events or dry periods.

Due to the insignificant difference of COD and ammonium removal between hydraulic regimes, the authors decided to continue operation from week 11 onwards with continuous flow only, since this is the regular regime for full-scale HF CWs. In addition, intermittently fed systems showed an extremely significant reduction in current density generation within the first transect (see Figure 3).

Average orthophosphate removal was very similar in the low OLR 1 period and slightly higher in continuously fed systems during high OLR period, however, without a statistically significant difference (for details see annex Table 5). A reason for the difference during high OLR period could be the temporarily (during feeding times) shortened HRT in intermittently fed systems leading to fewer orthophosphate removal through processes like adsorption and precipitation.

### 3.1.3 Organic loading effects

Overall, the removal efficiency of COD and ammonium did not depend on the OLR (low period one 4.9±1.6, high 13.6±3.2 g COD/$m^2$.day and low period two 6.7±1.4 g COD/$m^2$.day) and the thereby reduced HRT, showing no statistically significant differences (for details see annex Table 6). Total COD and ammonium removal on a mass basis was higher during the high OLR period, due to the higher influent concentrations (see Table 2). Despite the differing OLRs, removal rates in percentage showed that there were no real differences between OLR periods in COD or ammonium removal (see Table 4). In fact the removal efficiencies in percentage were rather increasing a little over time, from around 60% to 70% for COD and from around 25 to 40% for ammonium, probably due to the maturing of the systems. Both average nitrate and nitrite mass in closed-circuit systems increased during the high OLR period from in- to outlet. This could be interpreted as an effect of the observed increased ammonium removal through nitrification.

The systems adaptability to fluctuating organic loads illustrates a general asset of CWs; due to the fact that the majority of treatment happens in the first section of HF CWs, the remaining part of the system is able to lower the effects of flow and nutrient concentration peaks to a large degree, given that the systems are not overloaded or clogged (Samsó and García, 2014).

For the selection of the optimal OLR in CW-MFC systems it is important to find a good balance between the provision of sufficient substrate at the anode on the one side and overloading the system and thereby limiting the cathode functionality through growth of heterotrophic bacteria on the other (Doherty et al., 2015; Freguia et al., 2008; Villaseñor et al., 2013). Capodaglio et al. (2015) tested different OLRs in swine manure fed MFCs and found that lower OLR (volumetric OLR 0.7 kg COD/$m^3$.day) advantaged exoelectrogenic bacteria growth and activity over methanogenics as compared to higher OLR (volumetric OLR 11.2 kg COD/$m^3$.day). The highest OLR chosen in this study (corresponding to 0.06 kg COD/$m^3$.day) was governed by the given strength of the available urban wastewater and the highest hydraulic loading possible for continuous operation, given the size of the available feeding tank. Since the two tested OLRs in this study did not show significant differences, it seems they were within the above mentioned balanced range for the operation of CW-MFC systems, though rather on the very low end compared to MFC studies which used OLRs of a magnitude higher. However, OLRs in the presented study are in the range of conventional HF CW OLRs (Vymazal, 2005). Of course the OLR range for best performance is also dependent on the MFC architecture, e.g. the used anode with gravel and stainless steel mesh as electron acceptor has to be taken into account as well. Additionally, by offering a more favorable electron acceptor, MFCs have shown to postpone methane production, for example in experiments using plant MFCs (PMFC) inside rice microcosms (Arends et al., 2014) and in CW-MFCs (Fang et al., 2013).

With regards to electrical connections, although no significant differences were found within each of the three OLR periods, there was a slight tendency of increased treatment performance for closed-circuit systems in high OLR period and low OLR period 2. The authors believe that the absence of any difference among experimental conditions in continuously fed systems for the first experimental period (weeks 1-5) was due to the fact that the systems, and therefore the electrogenic biofilm, was still immature at the beginning of the experimentation, which is also reflected in the observed current, which was still increasing in all transects at the time (see Figure 3).

Low OLR 1 and high OLR periods had similar orthophosphate mass removal values although the influent load was doubled in the latter. Also, removal of orthophosphates in the last low OLR period 2 decreased below the levels of low OLR period 1 (see Table 2). These changes were probably not due to the different organic loading regimes but more likely due to the fact that phosphorus storage in CWs decreases over time due to finite capacity of adsorption sites in the biofilm and media (Kadlec and Wallace, 2009). In any case, the organic loading rate seems to have had no mentionable effect on orthophosphate removal in open- or closed-circuit systems.

### 3.1.4 Current

Figure 3 shows average current densities from the three MFCs corresponding to the three transects along the flow path for the intermittently and continuously fed closed-circuit systems.

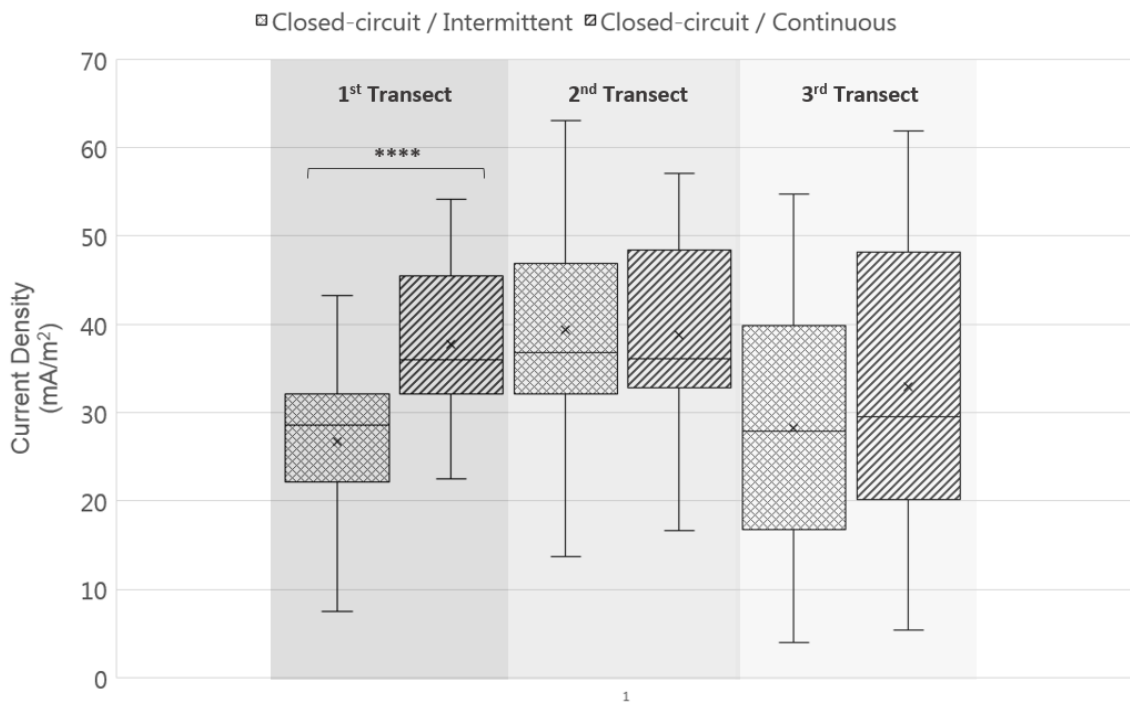

**** P-value < 0.0001

**Figure 3.** **Current density of intermittently and continuously fed closed-circuit systems per electrode and transect along the flow path during the first 10 weeks of experiments**

Average current densities (based on the projected anodic surface area) for closed-circuit/intermittent and closed-circuit/continuous systems per transect resulted in 26.8±9.4 and 37.7±8.1 mA/m$^2$ for the first electrode, 39.4±10.7 and 38.8±10.2 mA/m$^2$ for the second electrode and 28.2±9.4 and 32.9±17.1 mA/m$^2$ for the third electrode, respectively. Differences among hydraulic regimes were only statistically significant for the first transect ($p < 0.0001$) ($F_{(1, 68)}$; p = 3E-11), while differences in second ($F_{(1, 68)}$; p = 0.73) and third transect ($F_{(1, 68)}$; p = 0.08) were not significant.

Current results show that the hydraulic regime had an extremely significant effect on the first third of the systems with higher values in continuously fed systems.

With regards to OLR effect, Figure 4 shows the average current densities per transect of the four closed-circuit CW-MFC systems during different OLR periods interrupted by the summer break.

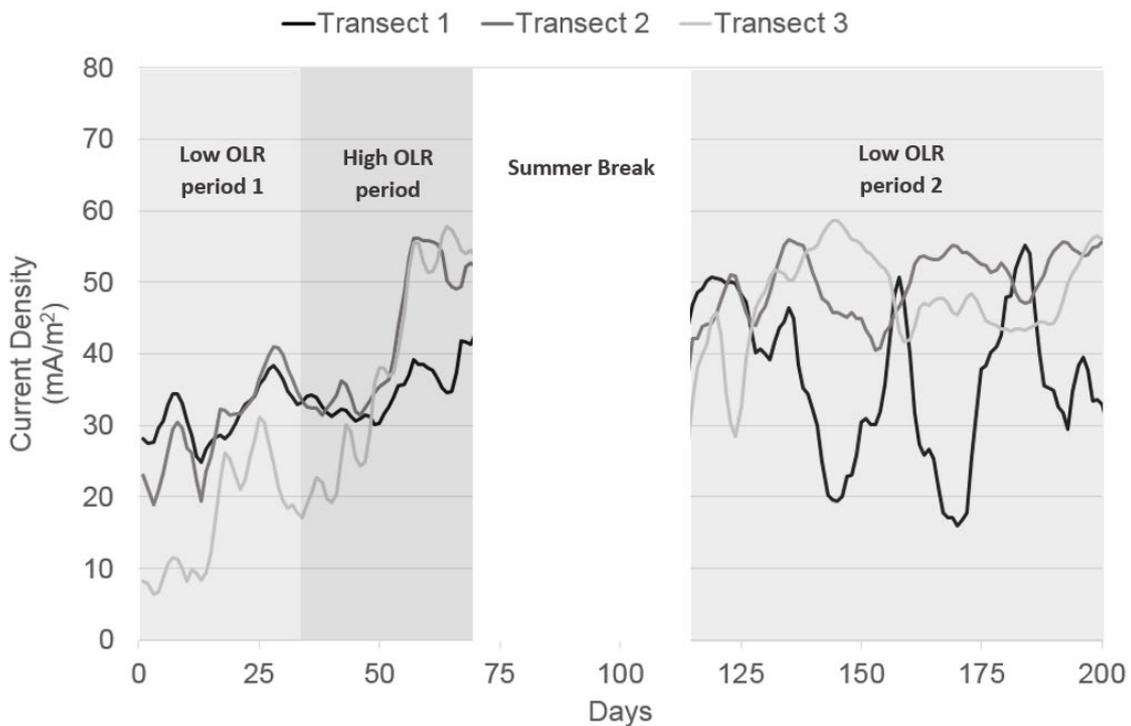

**Figure 4. Average current densities from four closed-circuit systems for each transect along time**

Current densities during low OLR period 1 were 33±6, 32±9 and 16±9 mA/m² for first, second and third transect, respectively. During the high OLR period current densities increased to 43±10, 45±11 and 43±13 mA/m² for first, second and third transect, respectively. Finally, during low OLR period 2 current densities amounted to 31±15, 49±9 and 50±7 mA/m² for first, second and third transect, respectively. Current densities in the first low OLR period were generally lower than in the following high and low OLR period 2. This is probably due to the incomplete maturity of the systems during the first weeks after experimentation start, rather than due to OLR effects, since current densities during the second low OLR period are of similar magnitude than those of the high OLR period.

### 3.2 Contaminant removal and microbial activity under different electrical connections

#### 3.2.1 Overview

In this section, contaminant removal efficiency of conventional, open-circuit and closed-circuit wetlands is addressed from the results obtained during week 12 to 23 of experimentation. During this period, all systems were operated in continuous flow with an average OLR of 6.7±1.4 g COD/m².

Table 3 summarizes the results of COD, ammonium, nitrate, nitrite and orthophosphate during the last 12 weeks of experimentation for all three electrical connections; CW control, open-circuit (OC) and closed-circuit (CC) CW-MFC systems. The results are shown as average mass at influent, after first transect, after second transect and effluent as well as removal from influent to effluent based on the average mass and percentage.

**Table 3.** Results for COD, ammonium, nitrate, nitrite and orthophosphate for CW control, open-circuit (OC) and closed-circuit (CC) CW-MFC systems during the last 12 experimentation weeks, expressed as average mass at influent, after first transect, after second transect and effluent as well as removal from influent to effluent based on the average mass and percentage.

|  |  | Influent | 1/3 | 2/3 | Effluent | Removal from Influent to Effluent | |
|---|---|---|---|---|---|---|---|
|  |  | (g/m$^2$.d) | | | | (g/m$^2$.d) | (%) |
| COD (n=11)[a] | CW | 6.6±1.5 | 3.3±1.0 | 2.5±0.6 | 2.0±1.1 | 4.5±1.0 | 69% |
|  | OC | 6.4±1.6 | 3.0±0.9 | 2.2±0.9 | 1.8±0.9 | 4.6±1.0 | 72% |
|  | CC | 6.7±1.5 | 2.9±1.0 | 2.1±0.9 | 1.7±0.9 | 4.9±1.1 | 74% |
| NH$_4$-N (n=7)[a] | CW | 1.2±0.2 | 1.1±0.2 | 0.9±0.2 | 1.0±0.3 | 0.3±0.3 | 19% |
|  | OC | 1.2±0.1 | 1.0±0.2 | 0.9±0.2 | 0.9±0.2 | 0.3±0.2 | 24% |
|  | CC | 1.3±0.1 | 1.0±0.1 | 0.8±0.2 | 0.7±0.2 | 0.5±0.3 | 41% |
| NO$_3$-N (n=8)[a] | CW | 0.002±0.007 | 0.000±0.000 | 0.0041±0.042 | 0.002±0.005 | 0.000±0.009 | -2% |
|  | OC | 0.001±0.004 | 0.000±0.000 | 0.031±0.023 | 0.001±0.004 | 0.000±0.000 | 0% |
|  | CC | 0.000±0.000 | 0.001±0.003 | 0.021±0.017 | 0.011±0.012 | -0.011±0.012 | NA[b] |
| NO$_2$-N (n=8)[a] | CW | 0.008±0.009 | 0.003±0.005 | 0.018±0.026 | 0.011±0.014 | -0.003±0.008 | -33% |
|  | OC | 0.011±0.017 | 0.014±0.017 | 0.034±0.017 | 0.015±0.019 | -0.004±0.014 | -40% |
|  | CC | 0.014±0.019 | 0.013±0.011 | 0.022±0.026 | 0.016±0.032 | -0.002±0.020 | -17% |
| PO$_4$-P (n=8)[a] | CW | 0.11±0.02 | 0.11±0.02 | 0.09±0.02 | 0.11±0.06 | 0.00±0.03 | 1% |
|  | OC | 0.11±0.02 | 0.10±0.02 | 0.09±0.02 | 0.09±0.02 | 0.01±0.01 | 10% |
|  | CC | 0.10±0.02 | 0.11±0.02 | 0.09±0.02 | 0.09±0.03 | 0.01±0.03 | 5% |

[a] Some experimentation weeks could not be considered due to highly diluted influent or technical analysis problems
[b] Division by zero

### 3.2.2 Electrical connection effect

As already previously described, closed-circuit systems on average outperformed open-circuit system during the first 10 weeks of operation (see Table 3), however, without significant differences (for details see annex Table 7). COD and ammonium removal from week 11 to 23 showed the same tendency but again without any significant difference. The same is true if compared with a CW control duplicate (from week 12 to 23) in the way that closed-circuit systems outperformed open-circuit and CW control systems as well, however, again without any significant difference. Again, the insignificance of differences, especially in the case of ammonium, could be partly due to the relatively high standard deviation most likely caused by the variation in quality of the used real urban wastewater due to natural causes like rainfall events or dry periods.

Average COD removal on a mass base in the last 12 weeks of experiments (the time when CW control was tested as well) in closed-circuit systems was only 2% higher than in open-circuit and 5% higher than in CW control systems (see Table 3). Wang et al. (2016b) found higher improvement with 8.3% difference in COD removal comparing closed- to open-circuit CW-MFC, however, using a pH control and vertically batch-fed bench-scale systems. Regardless the treatment around 75% of the overall COD mass removal was already removed within the first transect, between 15% and 20% in the second transect and between 5% and 10% in the last (see Figure 5).

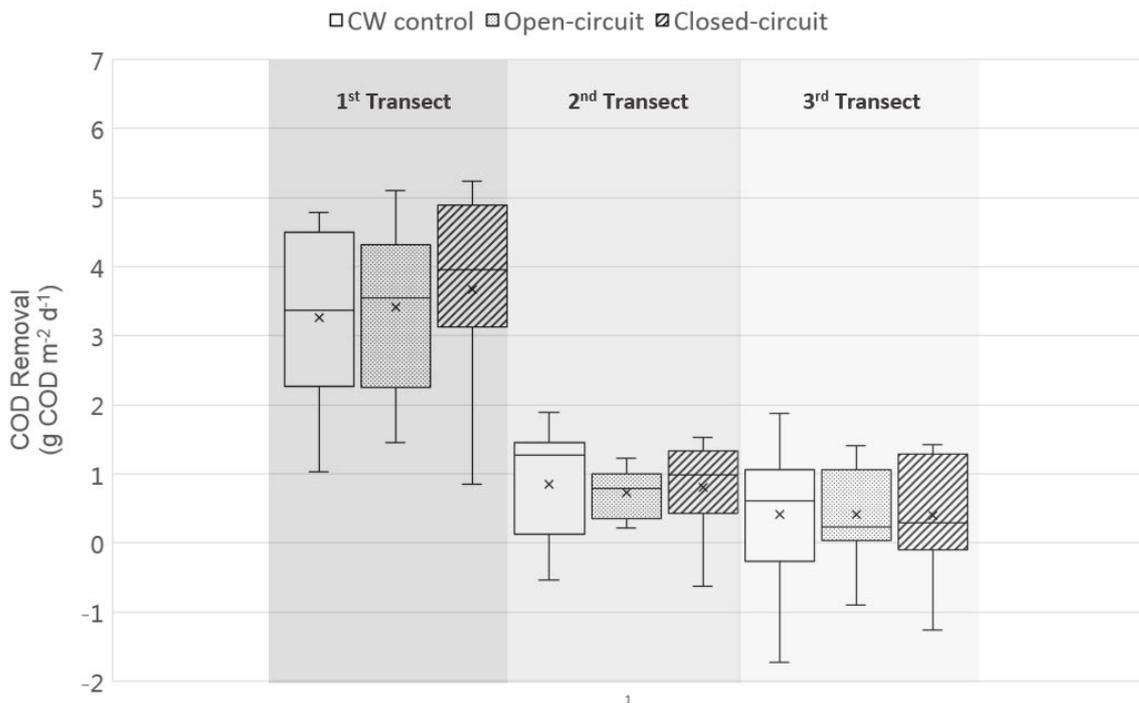

**Figure 5.** COD removal for each transect for CW control, open-circuit and closed-circuit systems (n=11, CW control duplicate started in week 12 and experimentation week 19 could not be used due to a highly diluted influent)

The overall COD removal of 74% in closed-circuit systems is comparable to earlier CW-MFC studies, with 75% (Yadav et al., 2012), 82% (Xu et al., 2018) and 76.5% (Zhao et al., 2013). In this regard, the presented study confirms results of these CW-MFC systems which were less representative for real situations; e.g. all mentioned above were in bench-scale, up-flow hydraulic regime, fed with synthetic or modified wastewater. Yadav et al. (2012) used very fine gravel (2-4 mm), only Xu et al. (2018) used a continuous flow but had a sand media and Zhao et al. (2013) used artificial aeration at the cathode. Some of these factors might influence treatment behavior, long term operation (e.g. clogging due to fine media) and possibly present up-scaling problems (e.g. flow direction, artificial wastewater). In comparison to full-scale HF CW systems the presented COD treatment efficiencies are not outstanding, but authors believe that the reason could be that meso- as well as lab-scale systems often have unfavorable hydraulic conditions due to the smaller scale, resulting in a lower HRT than the calculated theoretical HRT. An additional reason could be the lack of development of plants, which have shown to provide a significant positive wastewater treatment effect in subsurface flow CWs (Tanner, 2001).

Zhang et al. (2015) found indications through CE calculations in wastewater fed MFC systems (comparing closed- and open-circuit), that electrogenic bacteria outcompeted other microbial degradation pathways, while Fang et al. (2013) showed that electrogenic bacteria such as *Geobacter sulfurreducens* and *Beta Proteobacteria* inhibited the growth of *Archaea* at the anode. Although the difference in COD removal in the presented study is very low, the more competitive electroactive pathway and potential inhibition of non-electroactive bacteria could have been the reason for the increased COD removal in closed-circuit systems.

Average ammonium removal on a mass base in the last 12 weeks in closed-circuit systems was 17% higher than in open-circuit systems and 22% higher than in CW control (see Table 3) but not statistically different (for details see annex Table 7). Average ammonium removal in transects was not as homogeneous across treatments as for COD; in closed-circuit systems the majority was removed in the first and second transect and only a small portion in the last, in open-circuit systems the majority was removed in the first and the rest in even parts in second and third, and in CW control basically the whole treatment took place in the first and second transect (see Figure 6).

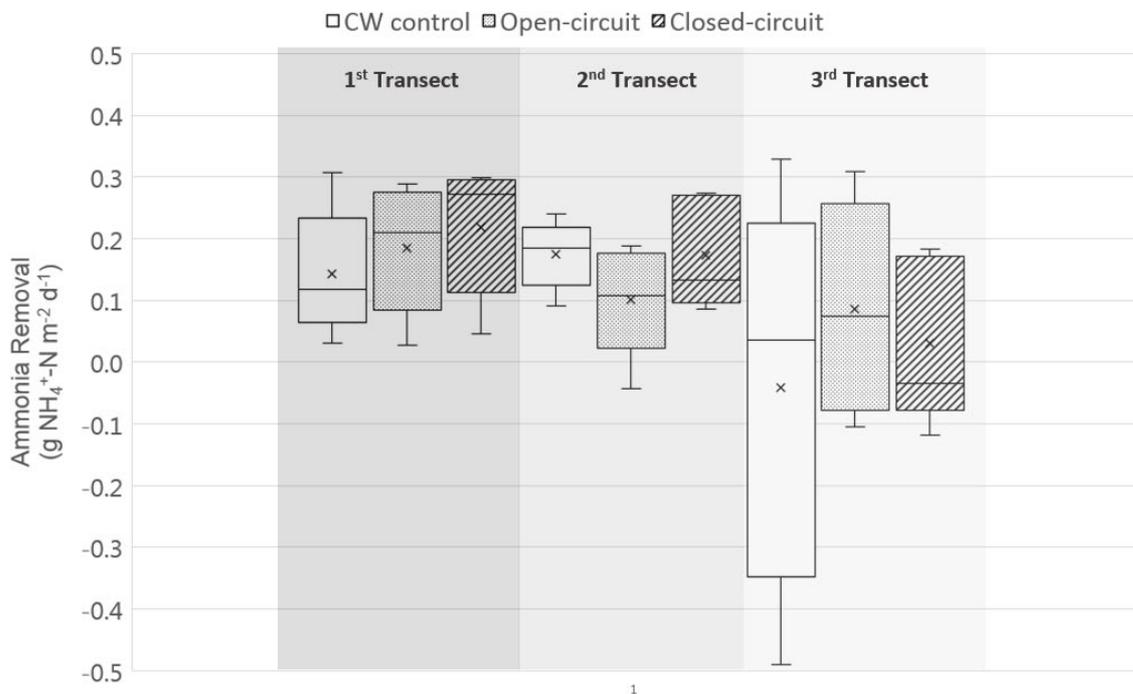

**Figure 6.** Ammonium removal per transect for CW control, open-circuit and closed-circuit systems (n=7; CW control duplicate started in week 12 and five experimentation weeks could not be used due to technical analysis or influent dilution problems due to rainfall)

The high variability in the last transect of CW control is remarkable and could indicate that it was more unstable than in open- or closed-circuit systems. Nitrate and nitrite effluent levels were generally very low during the time of electrical connections comparison (only week 11 was unusually high, but probably due to the start-up after summer). Both parameters increased a little in the second transect across all treatments and dropped again in the last (see Table 3). The only statistically significant difference between electrical connections occurred for nitrate when looking at the removal from inlet to outlet (for details see annex Table 7). Table 3 shows that the average nitrate level in closed-circuit CW-MFC systems was actually very similar after the first transect and even lower after the second transect as compared to CW control and open-circuit CW-MFC systems. Only in the last transect nitrate levels only dropped by nearly half in closed-circuit CW-MFC while they went close to the initial influent concentration in the other electrical connections.

The observed average ammonium removal of 41% in closed-circuit systems was rather low compared to preliminary results of Zhao et al. (2013) with an average of 77%, however, as mentioned above, the system had an artificially aerated cathode. In terms of improvement of efficiency compared to a control, Wang et al. (2016b) reported a 40% improvement of nitrate removal in closed-circuit CW-MFCs compared to open-circuit, however, with a pH control. Most other works on CW-MFCs were rather focused on organic matter and not on nitrogen removal. Xu et al. (2018) recently observed an average of 82% total nitrogen removal, however, the systems were continuously up-flow fed bench-scale systems with a tubular shape. Furthermore, Xu et al. (2018) did a functional analysis of the microbial community, comparing a closed-circuit CW-MFC with a CW control system, showing that (1) diversity and richness were higher in CW-MFC, (2) in the CW-MFC anode compartment the most common microbial functional groups were ammonia oxidizing bacteria (AOB), nitrite-oxidizing bacteria (NOB) and anaerobic ammonium oxidation (anammox) bacteria, with NOB and anammox being significantly higher than in the control and (3) in the CW-MFC cathode compartment the microbial functional groups denitrifying bacteria (DNB), dissimilatory nitrate reduction to ammonium (DNRA), and electroactive bacteria were significantly higher than in the control. In another microbial community analysis in CW-MFC systems, Wang et al. (2016b) found that anodes of closed-circuit as compared to open-circuit systems had a significantly improved richness in electroactive bacteria, nitrobacteria and DNB. Corbella et al. (2015) also found that *Geobacter* and methanogenic populations were significantly higher in closed-circuit when compared to open-circuit CW-MFC.

Of course the microbial community will also be dependent on the used materials for filter media, anode, cathode etc.; Wang et al. (2016a) found a significantly different distribution of microbial communities depending on the used CW-MFC anodes, comparing carbon fiber felt, graphite rods, foamed nickel and stainless steel mesh. Stainless steel mesh, the material used in this experiment, and foamed nickel had significantly lower relative abundance of *Proteobacteria* than carbon fiber felt and graphite rods, which was related to a lower power production. However, reported voltage outputs by Wang et al. (2016a) using stainless steel mesh reached averages from ca. 17 to 41 mV, which was by far surpassed in the presented systems with averages of 304±96, 462±33, and 457±50 mV for first, second and third transect, respectively.

The above described enrichment in anammox bacteria was already indicated in earlier research on MFC systems; Di Domenico et al. (2015) observed that MFC mode provides conditions favoring the cultivation of anammox in the anodic compartment of the anaerobic digestate fed systems used, without inoculating anammox bacteria at any point (only electroactive bacteria *G. sulfurreducens* were inoculated). In another bench-scale MFC experiment, Li et al. (2015), this time using synthetic wastewater, were able to prove higher abundance of anammox bacteria and associated higher nitrogen removal in closed-circuit MFC systems (open-circuit as control). However, these were inoculated with anammox bacteria in advance. Anammox bacteria were detected in conventional HF CW systems without MFC systems as well, however, Coban et al. (2015) could not detect any anammox activity in HF CWs inferring that the process is of low importance in the nitrogen removal of conventional CW systems.

Another possible ammonium removal pathway could be volatilization due to proton loss at the cathode and associated locally elevated pH, which cannot be excluded since the authors did not have the capability to measure pH on a micro-scale at the cathode, e.g. by using microprobes (Kim et al., 2008).

In MFC systems designed for nitrogen removal, simultaneous nitrification and denitrification (SND) could be accomplished; Virdis et al. (2008) observed that although oxygen was present at the cathode, biofilm stratification at the cathode allowed nitrifying bacteria in the outer layer and putative denitrifying bacteria were found in the inner layers in a micro-anoxic environment. However, large amounts of oxygen around the cathode would inhibit the bioelectrochemical denitrification (Kelly and He, 2014), which is the case for the presented systems, and again there would have been no possibility to measure SND in the presented experimental setup.

Conventional nitrification through supply with oxygen could have only happened at the systems very surface since DO measurements in the influent, effluent and the water column were always below detection limit, and therefore oxygen could have only partly been responsible for ammonium removal, which still could not have explained the differences between treatments. Xu et al. (2018) also described how, even in separator-less (e.g. without a membrane or glass-wool between anode and cathode) CW-MFCs, like the ones presented here, unwanted oxygen diffusion to the anode is inhibited by microorganisms which deplete the oxygen before it can reach further down, forming a so-called "microbial separator". This separator maintained also anaerobic conditions for the anode with just 2 cm distance from the cathode which showed the highest maximum power density compared to higher distances and systems with a separator. This distance is comparable to the distance between cathode and beginning of the anode (which extends vertically nearly until the bottom) in the presented work.

Orthophosphate removal during the first 10 weeks of operation differed only very slightly between treatments, again with higher rates in closed-circuit and continuously fed systems with a removal of up to 29% in closed-circuit continuous (see Table 3). Differences were not statistically significant (for details see annex Table 7). Ichihashi and Hirooka (2012) observed phosphate removal of 70-82% in closed-circuit MFC systems, with 4.6–27% in form of precipitation on the cathode, mainly in the form of struvite. While Corbella and Puigagut (2018) also found 15 % higher $PO_4^{-3}$ removal, comparing closed- to open-circuit CW-MFC systems, they also found white precipitation on the cathode which was not struvite but mostly Calcite ($CaCO_3$) and Halite (NaCl). However, maybe the conditions for struvite crystal precipitation were not met, i.e. $Mg^{2+}$, $NH_4$, and $PO_4^{-3}$ should exceed the solubility limit. Struvite solubility decreases with increasing pH (Doyle and Parsons, 2002). In addition, Zhang et al. (2012) found that biological phosphorus uptake, rather than chemical precipitation, can be increased in low current (smaller than 10 A) bioelectrochemical systems which is the case for the study of Corbella and Puigagut (2018) with ca. 1.45 mA and also the presented study with an average of ca. 1.48 mA across all three transects in the first 10 weeks. In any case, in the presented study no white precipitation was found on the cathodes.

Orthophosphate concentrations in the last 12 weeks basically stayed the same along the flow path across all three treatments. As described earlier it seems that adsorption sites already got limited in that period, since removal rates were higher in the first 10 weeks of experiments. In general, phosphorus storage in subsurface flow CWs takes place in plant biomass, bed media or accretion sediments and has a finite capacity (Kadlec and Wallace, 2009).

During the time of electrical connection comparison, from week 12 to 23, average voltages in the closed-circuit CW-MFC systems for the three transects amounted to 304±96, 462±33 and 457±50 V. Average current densities during the electrical connection comparison, from week 11 to 23, were 31±15, 49±9 and 50±7 $mA/m^2$ for transects 1, 2 and 3, respectively. These results are in the range of current densities in earlier CW-MFC experiments, with averages of 22.3 $mA/m^2$ by Villaseñor et al. (2013) and 70 $mA/m^2$ by Yadav et al. (2012). Polarization curves help to electrochemically characterize MFC systems and are shown for a closed-circuit CW-MFC replicate in the annex (see Figure 8). The resulting maximum power densities and corresponding current densities amounted to 6.7 mW/m$^2$ and 27.3 mA/m$^2$ in the first transect, 36.6 mW/m$^2$ and 92.8 mA/m$^2$ in the second transect and 35.9 mW/m$^2$ and 92.8 mA/m$^2$ in the third transect. The estimated internal resistances derived from the polarization curves were around 215 Ω, 100 Ω and 100 Ω for first, second and third transect, respectively. Principally, the potential maximum power is achieved when internal and external resistances are close to each other (Lefebvre et al., 2011). Therefore, it seems that the external resistance of 220 Ω fits very well for the first transect. According to the results, the second and third transect could potentially perform better with a lower external resistance around 100 Ω, however, it was decided to keep the same external resistance for all three transects for this experiment. The lower maximum power density in the first transect could be due to the higher organic loading in the first transect as compared to the second and third, which could a) potentially cause a clogging in the carbon felt cathode, limiting its potential and/or b) as also mentioned above in the discussion on the OLR, it was found that, in MFC systems, lower OLR benefited exoelectrogenic bacteria growth and activity over competing methanogenics (Capodaglio et al., 2015).

Coulombic Efficiency (CE) is the proportion of the produced charge to the carbohydrates which are theoretically derived from oxidation, indicated by the change of COD from transect to transect (Scott, 2016). The CEs over the whole time period in the three consecutive transects ranged from 0% to 8%, -34% to 46% and -89% to 93%, with averages of 1±3%, 10±17% and 2±34%, respectively. Earlier reported CW-MFC CEs range from 0.05-0.06% (Yadav et al., 2012) up to 2.8-3.9% (Liu et al., 2014). However, the authors believe that the parameter CE is not very useful for describing a CW-MFC's electric efficiency, especially if expressed per transect, since not only organic matter from the influent can contribute to the MFC signal but also accumulated organic matter within the gravel bed is a fuel source for MFC (Corbella et al., 2016a). This is probably the reason why the CE could reach high levels in the second and third transect; due to little COD removal and currents similar to the first transect it appears like a high current was produced with only little input. Therefore, the reported high positive CE values in this paper, especially in the second and third transect, are most likely overestimated. The second and third transect CE even reached negative values due to eventually increasing COD concentrations within the wetland caused by changes in influent wastewater quality.

### 3.2.3 Microbial activity

Figure 7 shows microbial activity, determined through the FDA experiment, along the flow path of the CW control systems and closed-circuit CW-MFC systems (all continuously fed).

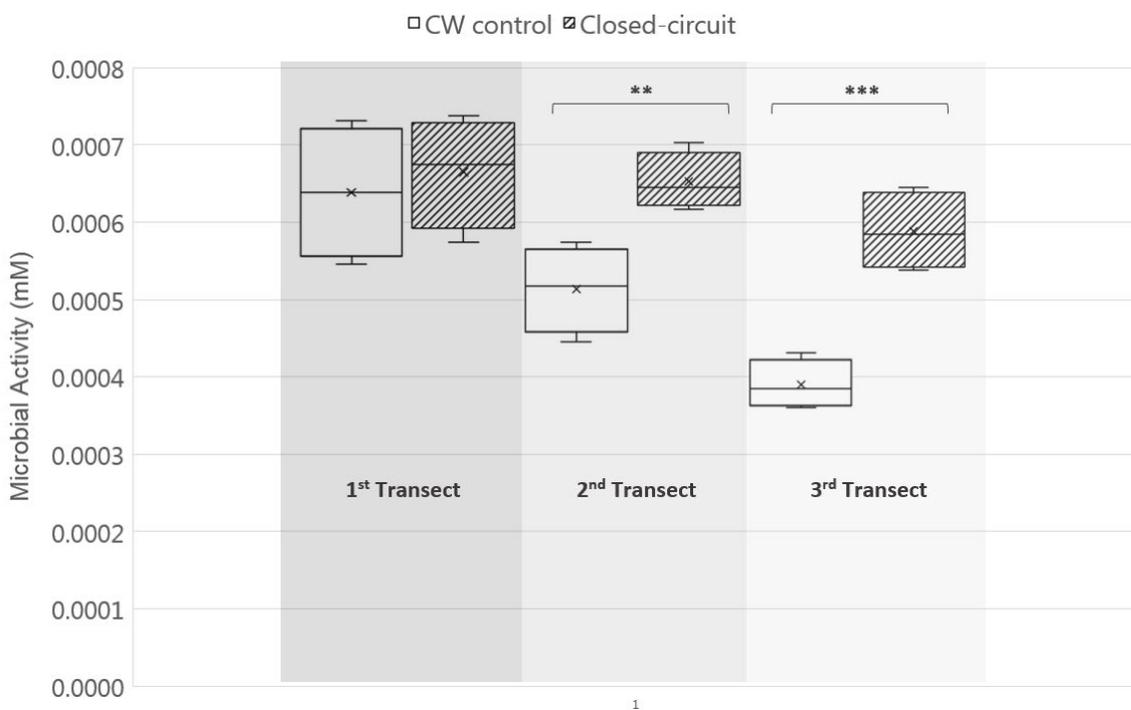

**Figure 7. Microbial activity along transects for control CW and closed-circuit continuously fed systems**

Generally, the activity was highest in the first transect, both in the closed-circuit and in CW control systems (activity analysis was not performed for open-circuit systems), and the activity stayed on a higher level in the closed-circuit as compared to the CW control systems. Differences between average microbial activities of closed-circuit and CW control systems were not statistically significant in the first transect (F (1, 4); p = 0.65), but statistically very significant in the second transect (p < 0.01) (F (1, 4); p = 0.006) and extremely significant in the third transect (p < 0.001) (F (1, 4); p = 0.0006).

The higher microbial activity within the first transect, irrespective of the treatment, is probably due to the higher availability of organic matter as a substrate, favoring the growth of microorganisms (Wu et al., 2014), with a subsequent decrease in microbial activity along the flow path, which has been observed already before in vertical and horizontal sequential CW systems (He et al., 2014). This decrease in activity is also reflected by the decrease in ammonium and COD removal along the systems flow path. Closed-circuit CW-MFC showed higher activity than CW control systems in all three transects. In percentages the microbial activity in closed-circuit systems was 4%, 21% and 34% higher than the control in first, second and third transect, respectively. Xu et al. (2018) analyzed diversity and richness (activity was not measured) of microbial communities in CW-MFC and CW control systems and found higher diversity and richness in closed-circuit CW-MFC systems. Also Wang et al. (2016b) found higher richness in closed-circuit as compared to open-circuit CW-MFC systems. Hence, in the presented systems a higher diversity and richness in closed-circuit CW-MFCs could have contributed to the measured higher activity. Corbella et al. (2015) also found that *Geobacter* and methanogenic populations were significantly higher in closed-circuit when compared to open-circuit CW-MFC.

As discussed in the section on COD removal comparing electrical connections, electrogenic bacteria in MFCs outcompeted other microbial communities and were also able to inhibit growth of *Archaea* at the anode (Fang et al., 2013; Zhang et al., 2015). This advantage in competition could be another factor responsible for the increased activity in the studied CW-MFC systems. Also, as mentioned above in the discussion on the OLR, it was found that, in MFC systems, lower OLR benefited exoelectrogenic bacteria growth and activity over competing methanogenics (Capodaglio et al., 2015). Therefore, a possible explanation for the varying differences in microbial activity between closed-circuit and CW control systems along the flow path could be that the decreasing OLR from transect to transect is leading from an insignificant difference in the first to a very significant difference in the second and extremely significant difference in the third transect. However, in comparison to the mentioned MFC studies, even the higher OLR at the influent of the presented study is already quite low (around a magnitude lower as in the MFCs), but in the range of OLRs in conventional HF CWs (Vymazal, 2005). Therefore, the presented results could give an indication that even a further decrease in OLR, from an already relatively low level, still causes a recognizable advantage to the exoelectrogenic over the methanogenic pathway.

MFCs have also been used for monitoring of microbial activity, in low contaminated environments like groundwater (Tront et al., 2008) or monitoring of anaerobic digestion processes (Liu et al., 2011).

## 4. CONCLUSIONS

The different tested organic loading rates and hydraulic regimes had no significant effect on treatment efficiency of COD or ammonium in the examined meso-scale horizontal-flow CW-MFC systems, but continuously fed systems showed slightly better treatment performance than intermittently fed systems. In addition, intermittent flow significantly decreased current production in the first transect of closed-circuit CW-MFC systems when compared to continuous flow.

In terms of electrical connection, closed-circuit CW-MFC systems were able to enhance treatment efficiency in comparison to open-circuit CW-MFC and CW control systems, however, again without significant differences, which might be due to the use of real urban wastewater which varied in strength over time due to natural causes like rainfall events or dry periods.

Microbial activity clearly decreased along the flow path, as did ammonium and especially COD removal. Microbial activity was higher in all three transects in closed-circuit mode when compared to control conditions, which could be one of the reasons for the observed enhancement of treatment performance. Differences between closed-circuit and control systems were not significant in the first transect but very significant in the second and extremely significant in the third, possibly indicating that the lower organic load along the flow path benefited the activity of electrogenic bacteria over competing non-electrogenic bacteria.

**ACKNOWLEDGEMENTS**

This project has received funding from the European Union´s Horizon 2020 research and innovation programme under the Marie Skłodowska-Curie grant agreement No 676070. This communication reflects only the authors' view and the Research Executive Agency of the EU is not responsible for any use that may be made of the information it contains. Marianna Garfí is grateful to the Spanish Ministry of Economy and Competitiveness (Plan Estatal de Investigación Científica y Técnica y de Innovación 2013-2016, Subprograma Ramón y Cajal (RYC) 2016).

**ANNEX**

**Table 4.** COD, ammonium and orthophosphate mass based average removal rate in percentage from inlet to outlet for low OLR 1, high OLR and low OLR 2 as well as intermittent or continuous flow hydraulic regime for open-circuit (OC) and closed-circuit (CC) CW-MFC systems.

| Removal (%) | | Low OLR 1 4.9±1.6 g COD/m².day | | High OLR 13.6±3.2 g COD/m².day | | Low OLR 2[a] 6.7±1.4 g COD/m².day |
|---|---|---|---|---|---|---|
| | | Intermittent flow | Continuous flow | Intermittent flow | Continuous flow | Continuous flow |
| COD | OC | 58% | 56% | 58% | 62% | 72% |
| (n=4/5/11)[b] | CC | 56% | 58% | 68% | 69% | 74% |
| $NH_4$ -N | OC | 23% | 29% | 18% | 30% | 24% |
| (n=4/5/7)[b] | CC | 27% | 38% | 35% | 39% | 41% |
| $NO_3$ -N | OC | -95% | -110 | 44 | -24 | 0% |
| (n=4/4/8)[b] | CC | -186 | -290 | -539 | NA^ | NA[c] |
| $NO_2$ -N | OC | 71% | 71% | 67% | -78% | -40% |
| (n=4/4/8)[b] | CC | 67% | 83% | 48% | -314% | -17% |
| $PO_4$ -P | OC | 21% | 29% | 10% | 11% | 10% |
| (n=4/4/8)[b] | CC | 21% | 29% | 10% | 16% | 5% |

[a] Low OLR 2 results are shown in more detail in the section 3.2 on the electrical connection effects
[b] Some experimentation weeks could not be considered due to highly diluted influent or technical analysis problems
[c] Division by zero

**Table 5.** Two-factor ANOVA (with replication) results for the comparison of the factors hydraulic regimes (intermittent vs. continuous) and electric connections (open-circuit vs. closed-circuit) as well as the interaction between the two factors, separated in low OLR 1 and high OLR periods.

| Two-factor ANOVA | | | p-value | | |
|---|---|---|---|---|---|
| | | | Hydraulic Regime | Electric Connection | Interaction |
| Low OLR 1 | COD | F (1, 4) | 0.94 | 0.93 | 0.87 |
| | $NH_4$ -N | F (1, 4) | 0.51 | 0.53 | 0.98 |
| | $NO_3$ -N | F (1, 4) | 0.67 | 0.64 | 0.75 |
| | $NO_2$ -N | F (1, 4) | 0.74 | 0.52 | 0.84 |
| | $PO_4$ -P | F (1, 4) | 0.66 | 0.85 | 0.86 |
| High OLR | COD | F (1, 5) | 0.45 | 0.96 | 0.94 |
| | $NH_4$ -N | F (1, 5) | 0.43 | 0.71 | 0.85 |
| | $NO_3$ -N | F (1, 4) | 0.0007 *** | 0.03 * | 0.10 |
| | $NO_2$ -N | F (1, 4) | 0.44 | 0.02 * | 0.78 |
| | $PO_4$ -P | F (1, 4) | 0.86 | 0.62 | 0.69 |

* significant difference (p < 0.05)
** very significant difference (p < 0.001)
*** extremely significant difference (p < 0.001)

**Table 6**. One-factor ANOVA (with replication) results for the comparison of low OLR 1 and high OLR periods (considering only continuously fed closed-circuit CW-MFC systems) based on removal percentages ($NO_3$ –N and $NO_2$ –N could not be calculated due to divison by zero)

| One-factor ANOVA | | p-value |
|---|---|---|
| COD | $F (1, 4)$ | 0.39 |
| $NH_4$ -N | $F (1, 4)$ | 0.84 |
| $PO_4$ -P | $F (1, 4)$ | 0.35 |

**Table 7.** One-factor ANOVA (with replication) results for the comparison of the electric connections during the low OLR 2 period, for the total system from inlet to outlet and each of the three transects separately.

| One-factor ANOVA | | p-value Electric Connection (low OLR 2 period) | | | |
|---|---|---|---|---|---|
| | | Inlet-Outlet | Transect 1 | Transect 2 | Transect 3 |
| COD | $F (2, 11)$ | 0.73 | 0.77 | 0.91 | 0.99 |
| $NH_4$ -N | $F (2, 7)$ | 0.16 | 0.55 | 0.29 | 0.67 |
| $NO_3$ -N | $F (2, 8)$ | 0.03* | 0.35 | 0.38 | 0.21 |
| $NO_2$ -N | $F (2, 8)$ | 0.74 | 0.33 | 0.73 | 0.71 |
| $PO_4$ -P | $F (2, 8)$ | 0.84 | 0.72 | 0.27 | 0.14 |

* significant difference ($p < 0.05$)

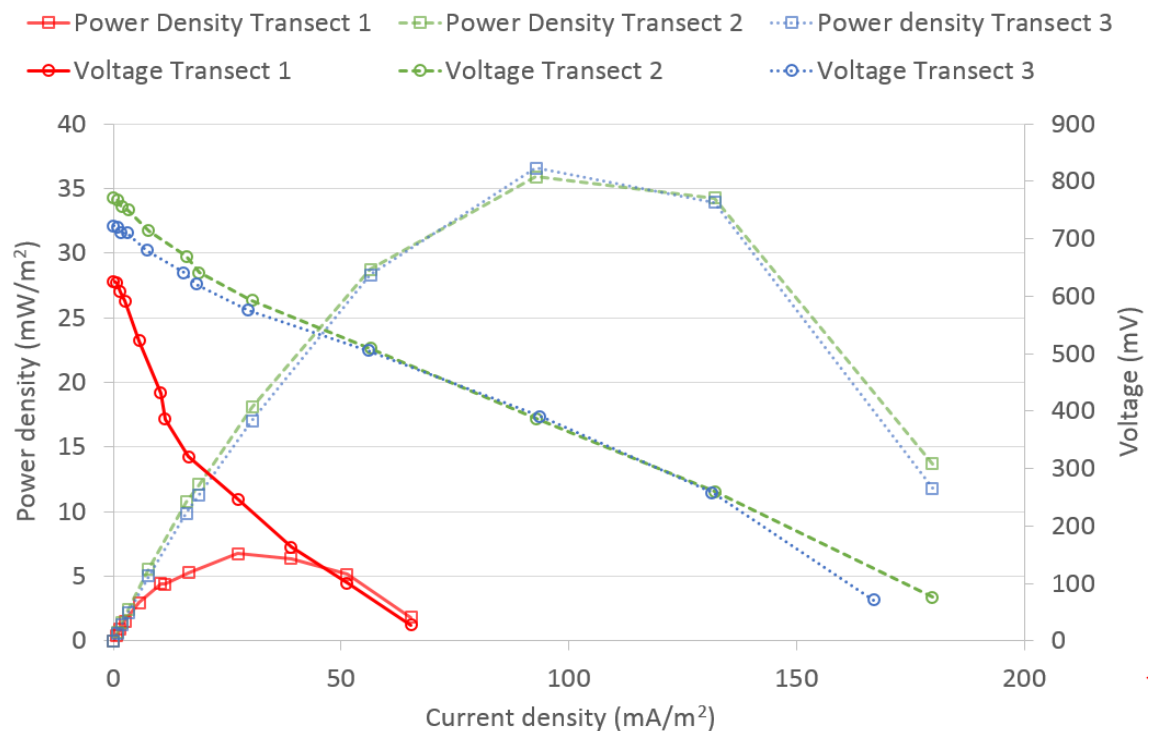

**Figure 8.** Power density and polarization curves for each transect of one of the closed-circuit CW-MFC replicates measured during sampling week 10